\documentclass[twocolumn,showpacs,preprintnumbers,amsmath,amssymb,superscriptaddress]{revtex4}

\usepackage{graphicx}
\usepackage{dcolumn}
\usepackage{bm}

\begin{document}


\title{The relaxation time of a chiral quantum R-L circuit}

\author{J. Gabelli}
\affiliation{Laboratoire Pierre Aigrain, D\'{e}partement de Physique
de l'Ecole Normale Sup\'erieure -
   24 rue Lhomond, 75005 Paris, France}
\author{G. F\`eve}
 \affiliation{Laboratoire Pierre Aigrain, D\'{e}partement de Physique de l'Ecole
Normale Sup\'erieure -
   24 rue Lhomond, 75005 Paris, France}
\author{T. Kontos}
 \affiliation{Laboratoire Pierre Aigrain,
D\'{e}partement de Physique de l'Ecole Normale Sup\'erieure -
   24 rue Lhomond, 75005 Paris, France}
  \author{J.-M. Berroir}
 \affiliation{Laboratoire Pierre Aigrain,
D\'{e}partement de Physique de l'Ecole Normale Sup\'erieure -
   24 rue Lhomond, 75005 Paris, France}
    \author{B. Placais}
   \email{Bernard.Placais@lpa.ens.fr}
 \affiliation{Laboratoire Pierre Aigrain,
D\'{e}partement de Physique de l'Ecole Normale Sup\'erieure -
   24 rue Lhomond, 75005 Paris, France}
\author{D.C. Glattli}
\affiliation {Laboratoire Pierre Aigrain, D\'{e}partement de
Physique de l'Ecole Normale Sup\'erieure - 24 rue Lhomond, 75005
Paris, France} \affiliation{Service de Physique de l'Etat
Condens\'{e}, CEA Saclay, F-91191 Gif-sur-Yvette, France.}

\author{B. Etienne}
\affiliation{Laboratoire de Photonique et Nanostructures, CNRS,
route de Nozay, F-91460 Marcoussis, France }
\author{Y. Jin}
 \affiliation{Laboratoire de Photonique et
Nanostructures, CNRS, route de Nozay, F-91460 Marcoussis, France }
\author{M. B\"uttiker}
 \affiliation{Universit\'e de Gen\`eve, 24 Quai Ernest Ansermet, CH-1211 Gen\`eve, Switzerland}

\date{\today}

\begin{abstract}
We report on the GHz complex admittance of a chiral one dimensional
ballistic conductor formed by edge states in the quantum Hall
regime. The circuit consists of a wide Hall bar (the inductor $L$)
in series with a tunable resistor ($R$) formed by a quantum point
contact. Electron interactions between edges are screened by a pair
of side gates. Conductance steps are observed on both real and
imaginary parts of the admittance. Remarkably, the phase of the
admittance is transmission-independent.  This shows that the
relaxation time of a chiral R-L circuit is resistance independent. A
current and charge conserving scattering theory is presented that
accounts for this observation with a relaxation time given by the
electronic transit time in the circuit.
\end{abstract}

\pacs{73.23.Ad,73.43.Cd,73.43.Fj,73.63.-b}
\maketitle

Violation of classical electro kinetic laws is a hallmark of quantum
transport. In the dc regime, it is well known that transport is
non-local over the electronic coherence length. This leads to the
non-additivity of parallel conductances \cite{Webb85PRL} and to
quantum composition laws to relate impurity scattering to
resistance.  Recently a similar manifestation of quantum coherence
has been reported by Gabelli et al.\cite{Gabelli06Science,note1} in
the ac regime where the resistance which determines the RC-charge
relaxation time of a mesoscopic capacitor is found to be quantized
at half of a resistance quantum. This observation, in agreement with
predictions of B\"uttiker, Thomas, and Pr\^etre
\cite{BPT93PRL,PTB96PRB}, establishes the concept of a charge
relaxation resistance \cite{note2} different from the standard dc
Landauer resistance. A second fundamental dynamical time scale is
the L/R-time of a mesoscopic circuit which in macroscopic conductors
is determined by the ratio of the inductance and the resistance of
the sample.

Here we investigate a series combination of an inductive and
resistive element and demonstrate that macroscopic kinetics does not
account for the correct ac response. In this case, chirality is
responsible for the observed non-classical behavior. The inductive
conductor is made of the {\it kinetic} inductance of electrons in
edge states \cite{halperin,buttiker,chklovskii} of a 2D electron gas
(2DEG) quantum Hall bar \cite{klitzing}. The resistive element is a
quantum point contact \cite{vanwees} (QPC). Theory
\cite{Christen96PRB} predicts that edge channels that connect two
reservoirs contribute to the impedance inductively due to kinetic
effects, whereas reflected edge channels contribute capacitively.
Importantly, in the present set-up, the inter-edge coupling is
reduced due to the large bar width and further minimized by using
side gates strongly coupled to the edge states. Our main result is
that the relaxation time of the quantum r-L circuit is not the
classical $L/R$ time but the electronic transit time of the circuit.

In our work the sample is still short compared to the wave length of
an edge-magneto-plasmon. Previous experimental investigations of the
electromagnetic response of Hall bars
\cite{Ashoori92PRB,talyanskii,Sukhodub2004PRL,Blick95APL} have
addressed the regime where the response is well accounted for by
collective excitations called edge-magneto-plasmons
\cite{Glattli85PRL} with wave-length short compared to the
dimensions of the sample. Refs.\cite{Ashoori92PRB,Sukhodub2004PRL}
have extensively studied the time domain and Ref.\cite{Blick95APL}
the frequency domain.

In this letter we report on phase-resolved impedance measurements of
a quantum R-L circuit in the edge state regime at GHz frequency and
milli-Kelvin temperatures. With increasing QPC transmission, dc-like
conductance steps are observed on both quadratures of the
admittance. Remarkably, the admittance phase is independent of the
number of transmitted modes and of their transmission. This shows
that the relaxation time of the chiral R-L circuit is resistance
independent. A current and charge conserving scattering theory
extending Ref.\cite{Christen96PRB} is presented that accounts for
this observation with a relaxation time given by the electronic
dwell time in the circuit.

\begin{figure}[ttt]
\includegraphics[scale=0.4]{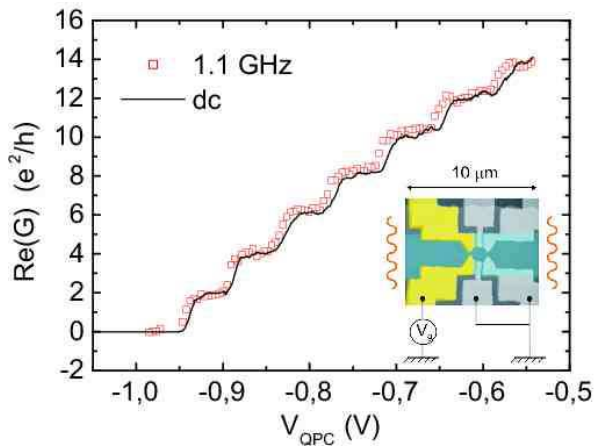}
\caption{\label{fig:epsart} Quantized steps in the dc conductance
and RF transmission of the circuit as function of the QPC gate
voltage. The solid line has been shifted by $+10mV$ along the
voltage axis to avoid curve overlapping which otherwise obscures
figure clarity. Temperature and magnetic field are respectively $50
mK$ and $0.224 T$. }\label{marches1}
\end{figure}

The sample is a  $50 \mu$m long and $6 \mu$m wide Hall bar made in a
GaAs/AlGaAs electron gas of nominal density $n_s=1.3\times 10^{11}$
cm$^{-2}$ and mobility $\mu=3\times 10^{6}$
cm$^{2}$V$^{-1}$s$^{-1}$. A magnetic field of $B=0.224 T$ and
$B=0.385 T$ is applied in the spin degenerate quantum Hall regime
(filling factors $N=24$ and $N=14$ respectively) so that edge states
are well developed. The bar is interrupted in its middle by a pair
of quantum point contacts (inset of Fig.\ref{marches1}). Only the
first QPC is active with a negative voltage bias ($V_g \sim -1$V).
Electron gas being fully depleted beneath it, the gate to 2DEG
capacitance is small. The grounded gate of the second QPC widely
overlaps the electron gas. This results in a large gate-2DEG
capacitance $c_g\sim 30$ fF (for a gate length $l_g\sim 10\mu$m)
which efficiently screens the inter-edge interactions. We estimate
$c_g\gg c_H$, with $c_H\sim 1$ fF the edge-to-edge capacitance for
the full Hall bar length. Long and wide non dissipative leads (not
shown in Fig.1) connect the sample to the contact pads.

The sample is mounted between two impedance-matched 50 $\Omega$
coplanar lines. Its impedance being large ($\gtrsim 10 k\Omega$),
the RF conductance $G(\omega)$ is simply proportional to the RF
transmission of the set-up. Phase is calibrated by assigning a
purely capacitive admittance ($\simeq40$fF) to the sample at the
pinch-off. This is corroborated by the vanishing of the dc
conductance.

Figure \ref{marches1} shows the real part $Re(G)$ at the opening of
the QPC. The large filling factor in the Hall bar ($N=24$) allows
the QPC to control the transmission of a large number of edge
states. As can be seen in the figure, the  RF data are proportional
to the dc one. In the following we shall assign the value $2e^2/h$
to the $Re(G)$ steps as a calibration of our set-up.

\begin{figure}[ttt]
\includegraphics[scale=0.7]{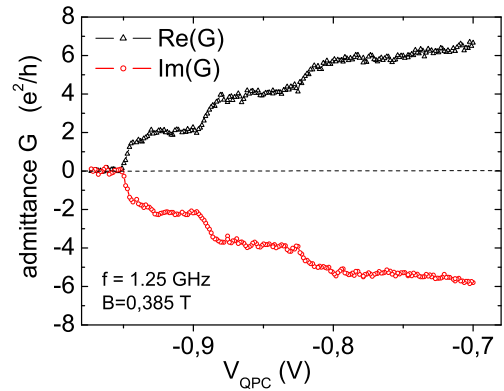}
\caption{\label{fig:epsart}Real and imaginary parts of the RF
admittance of the quantum Hall conductor as function of the QPC
voltage at $T=50mK$. Both signals show steps at the opening of the
first conducting channel. The negative imaginary part corresponds to
a negative emittance which is characteristic of an inductive
behavior. }\label{marches2}
\end{figure}

\begin{figure}
\includegraphics[scale=0.7]{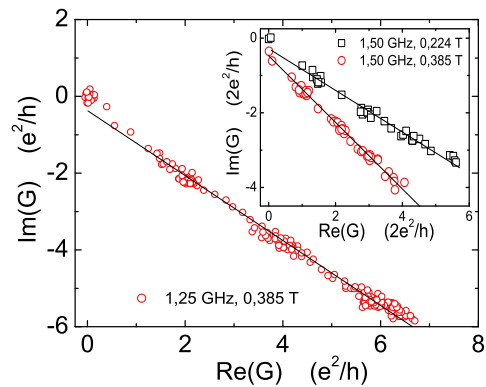}
\caption{\label{fig:epsart} Main frame: Nyquist representation of
the data of Fig.\ref{marches2}  showing that the admittance phase is
constant as function of the number of transmitted channels and of
their transmission. Point accumulation corresponds to the admittance
plateaus in Fig.\ref{marches2}. Insert : similar measurements
obtained at two different magnetic fields showing the linear
increase of the admittance phase with magnetic field.}
\label{Nyquist}
\end{figure}

Figure \ref{marches2} shows  $Re(G)$ and $Im(G)$ at $N=14$ for the
opening of the first three channels. Note that $Im(G)<0$ denotes an
inductive contribution. $Re(G)$ and $Im(G)$ show similar regular
steps as function of QPC transmission. The inductance step amplitude
is $\simeq 1\mu$H. In fact both quadratures are mutually
proportional as can be seen in the Nyquist plot of
Fig.\ref{Nyquist}. This corresponds to a transmission-independent
phase factor, $\tan(\varphi)=-\omega\tau=Im(G)/Re(G)$. It is well
explained by a constant relaxation time $\tau$, in strong contrast
with the time constant ($L/R \propto Re(G)$) of a classical circuit.
As an additional information, the inset depicts the linear magnetic
field dependence of $\tau$. These are the main results of our
experiment. We propose below an interpretation relying on the theory
developed by T. Christen and one of us for the low frequency
admittance of chiral conductors \cite{Christen96PRB}.

\begin{figure}[ttt]
\includegraphics[scale=0.25]{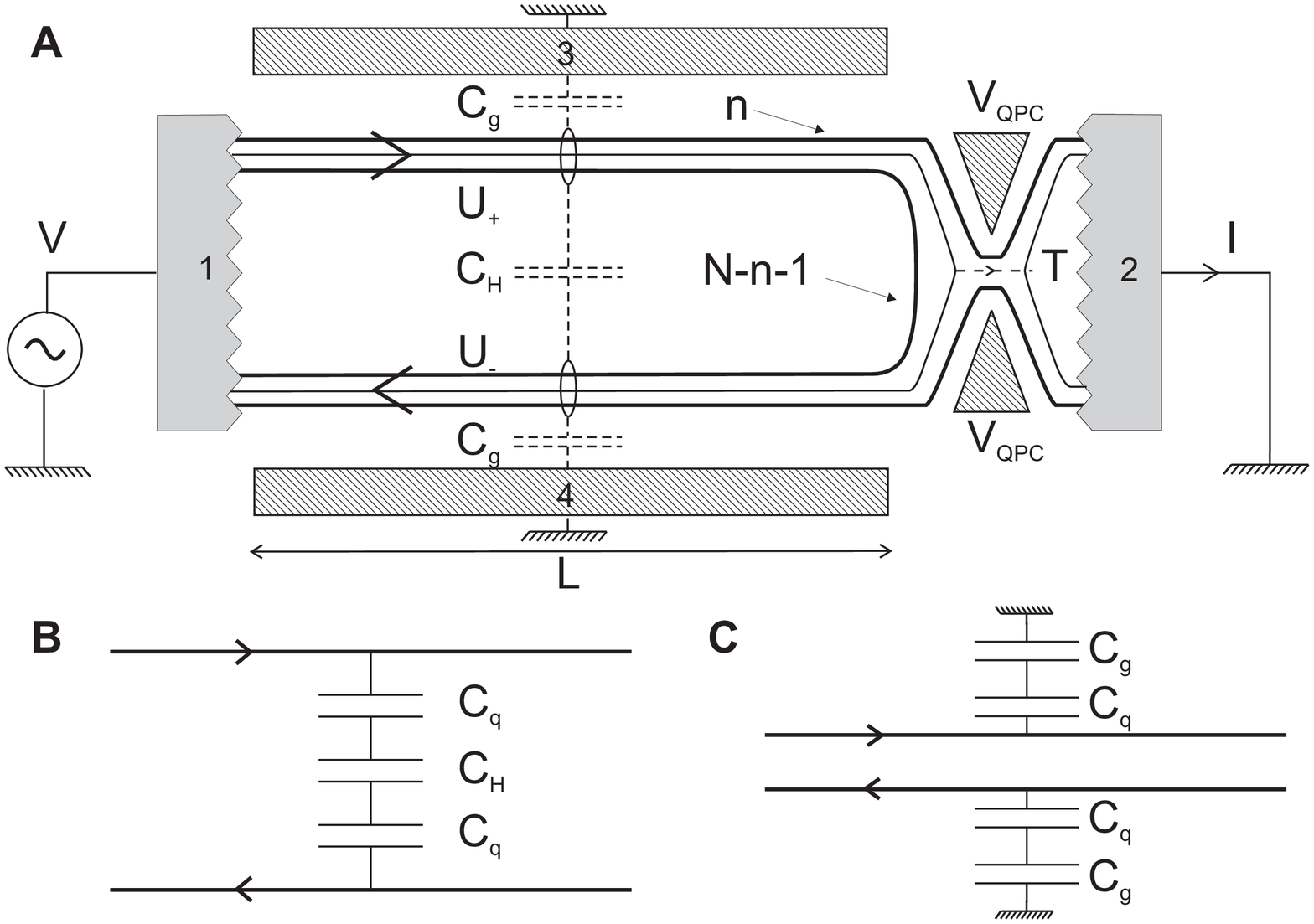}
\caption{\label{fig:epsart} A: Schematics of a quantum Hall bar with
$N$ edge states in series with a quantum point contact (QPC) with
$n$ fully transmitted channels and one partially transmitted
channel. Electrochemical equivalent circuit of a Hall bar in the
limit of weak edge to gate coupling (panel B) and weak interedge
coupling (panel C). Notations are specified in the text. }
\label{theoretical_skim}\end{figure}

In Ref.\cite{Christen96PRB} the emittance, $E=Im(G/\omega)$, has
been calculated for the case of a Hall bar with fully transmitted
and/or fully reflected edge states. The calculation takes into
account both inter-edge coupling and coupling to side gates. Here we
consider the case of a quantum Hall bar coupled to side gates in
series with a quantum point contact which controls the number of
transmitted channels and their transmission (see
fig.\ref{theoretical_skim}A). Let $N$ the number of filled Landau
levels (for simplicity we do not take into account spin degeneracy
in the calculations), $n$ be the number of fully transmitted modes
and $T$ the transmission of the partially transmitted one so that
(N-n-1) modes are totally reflected.  The length gate $l_g$ is small
enough that propagation effects can be neglected. Thus charging of
edge states is uniform but might differ on the upper and lower
branch of the edge state. Thus we can assume that the edge states on
the left upper side (labeled +) of the sample experience the same
electrostatic potential $U+$ and all the edge states on the left
lower side (labeled -) experience the potential $U-$. The upper and
lower edge states on the left side are equally coupled to side gates
with capacitance $c_g$ and the long range electrostatic interactions
between the upper and lower edges are described by a capacitance
$c_H$ (see fig.1.B). For simplicity, we take all left edge states to
have the same density of states, $\nu=l_g/hv_D$, where $v_D$ is the
drift velocity. $v_D$ is the ratio of the confining electric field
to the applied magnetic field and is therefore $\propto N$. The
quantum capacitance per channel is given by $c_q=e^2 \nu=l_g
e^2/hv_D$.

The low frequency response of the conductor is of the form
$dI_{\alpha}(\omega)/dV_{\beta} (\omega)= G_{\alpha \beta} - i
\omega E_{\alpha\beta} +..$ where $\alpha,\beta$ label current
contact indices $1$ and $2$ and gate indices $3$ and $4$
\cite{note4}. $G_{\alpha \beta}\neq0$ only for current contacts.
$E_{\alpha\beta}$ is a four terminal emittance matrix for the
quantum conductor with its gates.  According to
Ref.\cite{Christen96PRB}, the emittance is
\begin{equation}
E_{\alpha\beta}=e^2\sum_{k=+,-}\left[\frac{dN_{\alpha
k\beta}}{dE}-\frac{dN_{\alpha k}}{dE}u_{k,\beta}\right]\quad ,
\label{equ1}
\end{equation}
where $\frac{dN_{\alpha k\beta}}{dE}$ is the partial density of
states of carriers injected in contact $\beta$ that reach the upper
edge $k=+$, (or the lower edge $k=-$), and exit the sample through
contact $\alpha$. $\frac{dN_{\alpha k}}{dE}=\Sigma_\beta
\frac{dN_{\alpha k\beta}}{dE}$ is the emissivity of region $k=\pm$
irrespective of the contact from which the carriers are incident.
The characteristic potential $u_{k\beta}$ relates the change of the
electrostatic potential of conductor $k$ to that of the
electrochemical potential of contact $\beta$. In our geometry the
only non-zero partial density of states are:
\begin{eqnarray*}
\frac{dN_{1,\pm,1}}{dE}&=& ((1-T)+N-(n+1))\nu\\
\frac{dN_{2,+,1}}{dE}&=&\frac{dN_{1,-,2}}{dE}= (T+n)\nu
\label{equ3}\end{eqnarray*} from which we find the emissivities
$\frac{dN_{\alpha k}}{dE}$. We next need to find the characteristic
potentials $u_{k,\beta}$ on the upper and lower edges of the
conductor $k = \pm$ for each of the four contacts $\beta = 1, 2, 3,
4$. To this end we follow closely Ref.\cite{Christen96PRB} and find
the emittance matrix. The two-terminal admittance measurement
considered in this work is determined by the matrix element
\cite{note3} $E_{2,1} \equiv E$. We find
\begin{equation}
E= -c_{\mu g} \frac{(T+n)}{N} - \eta c_{\mu H} \frac{(T+n)^2}{N^2}
\quad  \label{equ8}\end{equation}\ where
\begin{eqnarray}
c_{\mu g}=\frac{Nc_q \, c_g}{c_g + N c_q}\quad &,& \quad c_{\mu
H}=\frac{Nc_q \, c_H}{2c_H + Nc_q}\label{toto} \label{equ9a}\\
\eta=\frac{1}{1+c_g/Nc_q}&\times&\frac{1}{1+c_g/(2c_H+Nc_q)}
\label{equ9b}\end{eqnarray} are respectively the electrochemical
capacitance between one edge and its side-gate and the mutual
capacitance of the edge states across the Hall bar. The coefficient
$\eta<1$ vanishes for strong gate coupling ($c_g\gg c_q$). Note that
$v_D\propto N$, so that $N\nu$, $c_{\mu H}$ and $ c_{\mu g}$ do not
depend on $N$.

For our experiment where inter-edge coupling is weak
($c_H<<c_q\lesssim c_g$), $E\simeq -c_{\mu g} (T+n)/N$,  we obtain
\begin{equation}
G(\omega)=G_0\left(1-i\omega\frac{h}{e^2}\frac{c_{\mu
g}}{N}\right)\, \label{equ13}\end{equation} where $G_0=(n+T)e^2/h$
is the Landauer dc conductance. Remarkably $G(\omega)$ exhibits a
transmission independent phase in agreement with the experiment.
Here the negative bar emittance can be interpreted with the
equivalent circuit in Fig.4C in terms of leakage currents to the
gate. It can be shown that the classical addition of the Hall bar
 and the QPC impedances gives a different result with a transmission-dependent
phase factor. This corresponds to the situation where a fictitious
reservoir is inserted between the two components which amounts
essentially to break the chirality of the experiment. We thus
observe a violation of classical laws which is here a pure effect of
chirality. The phase factor is given by the transit time of
electrons through the Hall bar $\tau=l_g/\widetilde{v_D}$, where
$\widetilde{v_D}= v_D+\frac{e^2Nl_g}{h c_g}$ is the drift velocity
which takes into account the screening by the side gate. Note that
$\widetilde{v_D}\propto N$ is inversely proportional to the magnetic
field. This magnetic field dependence is clearly observed in the
experiment (inset of Fig.\ref{Nyquist}). From the slope
($\simeq-0,89$) of the Nyquist diagram obtained at $B=0.385T$ (main
frame), and $l_g\sim10\mu$m , we estimate a drift velocity of
$\widetilde{v_D}\sim 10^5$ m/s in order of magnitude agreement with
the numbers in the literature \cite{Sukhodub2004PRL}.

We consider now the opposite case of a non-chiral quantum wire
(\emph{i.e.} a carbon nanotube) with strong inter-edge coupling
($c_g<<c_q\lesssim c_H$). We obtain
\begin{equation}
G(\omega)=G_0\left(1-i\omega G_0\frac{h^2}{e^4}c_{\mu H}\right).
\label{equ10}\end{equation} It corresponds to the lowest order
development for the admittance of the Landauer resistance $1/G_0$ in
series with the Hall-bar electrochemical inductance
$L_\mu=\frac{h^2}{e^4}c_{\mu H}/N^2$. In this non-chiral situation
classical laws are recovered. The negative Hall bar emittance can be
interpreted with the equivalent circuit in
Fig.\ref{theoretical_skim}B, in terms of displacement
counter-currents proportional to frequency. For very strong
inter-edge coupling ($c_{H}\rightarrow\infty$), $L_\mu$ reduces to
the usual kinetic inductance of a quantum wire,
$L_{kin}=\frac{h^2}{e^2}\frac{\nu}{2N}$
\cite{Burke2000APL,Burke02IEEE}. The signature of this regime would
be a linear variation of the admittance phase (\emph{i.e.} $\tau$)
with transmission, which corresponds to a circle arch in the Nyquist
diagram.

In conclusion, we have provided phase resolved measurements of the
admittance of a quantum Hall bar coupled to gates in series with a
quantum point contact. This realizes the simplest {\it chiral}
quantum R-L circuit. We observe quantized steps in both the active
and reactive parts of the admittance with a remarkable
transmission-independent phase. The phase is directly related to the
transit time of the electrons in the Hall bar. This interpretation
is further supported by the expected magnetic field dependence of
the transit time.
Our measurements are well described by a scattering
theory in the limit of strong side-gate coupling, allowing for a
direct determination of the electronic transit-time.
Our work demonstrates that interesting novel transport quantities such as
the mesoscopic analogs of the RC and L/R-times become accessible in the GHz
range provided the measurement is carried out on a sample with properties
that can be tuned over a wide range for instance as here with a QPC.

\begin{acknowledgments}

The Laboratoire Pierre Aigrain (LPA) is the CNRS-ENS UMR8551
associated with universities Paris 6 and Paris 7. The research has
been supported by ANR-05-NANO-028. The work of MB was supported by
the LPA, the Swiss NSF and the STREP project SUBTLE.

\end{acknowledgments}


\begin{thebibliography}{0}
\bibitem{note4}
To map the dynamics onto L-R circuit elements considered in this
work a first order expansion is sufficient.
\end{thebibliography}
\end{document}